\documentclass[11pt,twocolumn]{article}
\usepackage{amsfonts,amsmath,amssymb}
\newcommand{\Rl}{\mathbb{R}}

\newcommand{\Ir}{\mathbb{Z}}
\newcommand{\Cx}{\mathbb{C}}
\newcommand{\A}{\mathcal{A}}

\renewcommand{\H}{\mathcal{H}}

\newcommand{\bS}{\vec{S}}
\newcommand{\Tr}{{\rm Tr}\, }
\newcommand{\ran}{{\rm ran}\, }

\newcommand{\eq}[1]{(\ref{#1})}
\def\QED{{\hspace*{\fill}{$\square$}}\quad
          \vspace{10pt}}

\def\idty{{\mathchoice {\rm 1\mskip-4mu l} {\rm 1\mskip-4mu l} %
{\rm 1\mskip-4.5mu l} {\rm 1\mskip-5mu l}}}
\newcommand{\be}{\begin{equation}}
\newcommand{\ee}{\end{equation}}
\newcommand{\bea}{\begin{eqnarray}}
\newcommand{\eea}{\end{eqnarray}}
\newcommand{\beann}{\begin{eqnarray*}}
\newcommand{\eeann}{\end{eqnarray*}}
\DeclareMathAlphabet{\mathol}{OT1}{cmr}{l}{ol}

\newcommand{\Hil}{\mathcal{H}}

\begin{document}


\begin{center}
{\Large \bf Quantum Spin Systems\footnote{\copyright\ 2004 
Bruno Nachtergaele.}\\[27pt]}
{\bf Bruno Nachtergaele\footnote{
Supported in part by the National Science Foundation 
under Grant \# DMS-0303316.}}\\
Department of Mathematics\\
University of California, Davis\\
Davis, CA 95616-8633, USA\\[5pt]
{bxn@math.ucdavis.edu}\\[15pt]
\end{center}

\section{Introduction}

The theory of quantum spin systems is concerned with properties of quantum
systems with an infinite number of degrees of freedom that each have a
finite-dimensional state space. Occasionaly, one is specifically interested in
finite systems. In the most common examples one has an $n$-dimensional Hilbert
space associated with each site of a $d$-dimensional lattice.

A model is normally defined by describing a Hamiltonian or a family of
Hamiltonians, which are self-adjoint operators on the Hilbert space, and one
studies their spectrum, the eigenstates, the equilibrium states, its dynamics,
non-equilibrium stationary states etc.

More particularly, the term ``quantum spin system'' often refers to such models
where each degree of freedom is thought of as a spin variable, i.e., there
are three basic observables representing the components of the spin, $S^1,
S^2$, and $S^3$, and these components transform according to a unitary 
representation of $SU(2)$. The most commonly encountered situation is where 
the system consists of $N$ spins, each associated with a fixed irreducible
representation of $SU(2)$. One speaks of a spin$-J$ model, if this
representation is the $2J+1$-dimensional one. The possible values
of $J$ are $1/2,1,3/2,\ldots$.

The spins are usually thought of as each being associated with a site in a
lattice, or more generally, a vertex in a graph. E.g., each spin may be
associated with an ion in a crystaline lattice, which is how quantum spin
models arise in condensed matter physics. Quantum spin systems are also used
in quantum information theory and quantum computation, and show up as
abstract mathematical objects in representation theory and quantum
probability.

In this article we give a short introduction to the subject, starting
with a very brief review of its history. In Section \ref{sec:frame} we sketch
the mathematical framework and give the most important definitions. Three
further sections are entitled {\em Symmetries and symmetry breaking}, {\em
Phase transitions}, and {\em Dynamics}, which together cover the most
important aspects of quantum spin systems actively pursued today.

\section{A very brief history}

The introduction of quantum spin systems was the result of the marriage of two
developments taking place in the 1920's. The first was the
realization that angular momentum (hence, also the magnetic moment) was
quantized (Pauli, 1920; Stern and Gerlach, 1922) and that particles such as the
electron have an intrinsic angular momentum called spin (Compton, 1921;
Goudsmit and Uhlenbeck, 1925). 

The second development was the attempt in statistical mechanics to explain
ferromagnetism and the phase transition associated with it on the  basis of a
microscopic theory (Lenz and Ising, 1925). The fundamental interaction  between
spins, the so-called {\em exchange} operator which is a subtle consequence of
the Pauli exclusion principle, was introduced independently by Dirac and 
Heisenberg in 1926. With this discovery it was realized that magnetism is a
quantum effect and that a fundamental theory of magnetism requires the study of
quantum mechanical models. This realization and a large amount of subsequent
work notwithstanding, some of the most fundamental questions, such as a
derivation of  ferromagnetism from first principles, remain open.

Heisenberg gave his name to the first and most important quantum spin model,
the {\em Heisenberg model} (see further). It has been studied intensely ever
since the early 1930's and its study has led to an impressive variety of new
ideas in both mathematics and physics. Here, we limit ourselves to listing some
landmark developments.

{\em Spin waves} were discovered independently by Bloch and Slater in 1930.
Spin waves continue to play an essential role in our understanding of the
excitation spectrum of quantum spin Hamiltonians. In two papers published in
1956, Dyson advanced the theory of spin waves by showing how interactions
between spin waves can be taken into account.

In 1931, Bethe introduced the famous {\em Bethe Anstaz} to show how the exact
eigenvectors of the spin 1/2 Heisenberg model on the one-dimensional lattice
can be found. This exact solution, directly and indirectly led to many important
developments in statistical mechanics, combinatorics, representation theory,
quantum field theory and more. Hulth\'en used Bethe's Ansatz to compute the ground state energy
of the antiferromagnetic spin-1/2 Heisenberg chain in 1938.

In their famous 1961 paper Lieb, Schultz, and Mattis showed that some quantum
spin models in one dimension can be solved exactly by mapping them into a
problem of free Fermions. This paper is still one of the most cited in the
field.

Robinson, in 1967, laid the foundation for the mathematical framework that we 
describe in the next section. Using that framework, Araki established the
absence of phase transitions at positive temperature a large class of
one-dimensional quantum spin models in 1969.

During the more recent decades the mathematical and computational techniques
used to study quantum spin models have fanned out in many directions.

When it was realized in the 1980's that the magnetic properties of complex
materials play an important role in  high-$T_c$ superductivity, the variety of
quantum spin models studied in the literature exploded. This motivated a large
number of theoretical and experimental studies of materials with exotic
properties that are often based on quantum effects that do not have a classical
analogue. An example of unexpected behavior is the prediction by Haldane of the
spin liquid ground state of the spin-1 Heisenberg antiferromagnetic chain in
1983. In the quest for a mathematical proof of this prediction (a quest still
ongoing today), Affleck, Kennedy, Lieb, and Tasaki introduced the AKLT model in
1987. They were able to prove that the ground state of this model has all the
characteristic properties predicted by Haldane for the Heisenberg chain: a
unique ground state with exponential decay of correlations and a spectral gap
above the ground state.

There also are particle models that are defined on a lattice, or more
generally, a graph. Unlike spins, particles can hop from one site to another.
These models are closely related to quantum spin systems and in some cases
mathematically equivalent. The best known example of a model of lattice
fermions is the Hubbard model. We will not further discuss such systems in this
article.

\section{Mathematical Framework}\label{sec:frame}

Quantum spin systems is an area of mathematical physics where the demands of
mathematical rigor can be fully met and in many cases this can be done
without sacrificing  the ability to include all physically relevant models and
phenomena. This does not mean, however, that there are few open problems
remaining.  But it does mean that, in general, these open problems are
precisely formulated mathematical questions. 

In this section, we will review the standard mathematical framework for
quantum spin systems, in which the topics discussed in the subsequent section
can be given a precise mathematical formulation. It is possible, however, to
skip this section and read the rest with only a physical or intuitive
understanding of the notions of observable, Hamiltonian, dynamics, symmetry,
ground state etc...

\newcommand{\fsets}{\mathcal{L}}

The most common mathematical setup is as follows. Let $d\geq 1$, and let
$\fsets$ denote the family of finite subsets of the $d$-dimensional integer
lattice $\Ir^d$. For simplicity we will assume that the Hilbert space of the
``spin'' associated with each $x\in\Ir^d$ has the same dimension $n\geq 2$:
$\Hil_{\{x\}}\cong\Cx^n$. The Hilbert  space associated with the finite volume
$\Lambda\in \fsets$ is then  $\Hil_\Lambda = \bigotimes_{x\in \Lambda}
\Hil_x$. The algebra of observables for the spin of site $x$ consists of the
$n\times n$ complex matrices: $\A_{\{x\}}\cong M_n(\Cx)$. For any $\Lambda\in
\fsets$, the algebra of observables for the system in $\Lambda$ is given by
$\A_\Lambda =\bigotimes_{x\in\Lambda} \A_{\{x\}}$. The primary observables for
a quantum spin model are the spin-$S$ matrices $S^1, S^2$, and $S^3$, where
$S$ is the half-integer such that $n=2S+1$. 
They are defined by the property that they are Hermitian
matrices satisfying the $SU(2)$ commutation relations.
Instead of $S^1$ and $S^2$, one often works with the spin raising and lowering
operators, $S^+$ and $S^-$, defined by the relations 
$S^1 = (S^+ + S^-)/2$, and $S^2 = (S^+ - S^-)/(2i)$.
In terms of these, the $SU(2)$ commutation relations are
\be
[ S^+ , S^- ] = 2 S^3, \quad
[ S^3 , S^\pm ] = \pm S^\pm \, ,
\label{commutation}\ee
where we have used the standard notation for the commutator for two elements
$A$ and $B$  in an algebra: $[A,B]=AB-BA$.
In the standard basis $S^3, S^+$, and $S^-$ are given by the following matrices:
$$
S^3 = \left(
  \begin{array}{cccc}S \\ & S-1 \\ & & \ddots \\ & & & -S\end{array}
  \right)\, ,
$$
$S^- = (S^+)^*$, and
$$
S^+ = \left(
  \begin{array}{ccccc}0 & c_S \\ & 0 & c_{S-1} \\ & & \ddots & \ddots 
\\ & & & 0 & c_{-S+1} \\ & & & & 0\end{array} \right)\, ,
$$
where, for $m=-S,-S+1,\ldots, S$,
$$
c_m = \sqrt{S(S+1) - m(m-1)} \, .
$$
In the case $n=2$, one often works with the {\em Pauli matrices},
$\sigma^1,\sigma^2,\sigma^3$, simply related to the spin matrices by 
$\sigma^j=2S^j$, $j=1,2,3$.

Most physical observables are expressed as finite sums and products 
of the spin matrices $S^j_x$, $j=1,2,3$, associated with the site $x\in\Lambda$:
$$
S^j_x = \bigotimes_{y\in\Lambda} A_y
$$
with $A_x=S^j$, and $A_y=\idty$ if $y\neq x$.

The $\A_\Lambda$ are finite-dimensional 
$C^*$-algebras for the usual operations of sum, product, and Hermitian
conjugation of matrices and with identity $\idty_\Lambda$.

If $\Lambda_0 \subset \Lambda_1$, there is a natural embedding of
$\A_{\Lambda_0}$ into $\A_{\Lambda_1}$., given by 
$$
\A_{\Lambda_0}\cong\A_{\Lambda_0}\otimes \idty_{\Lambda_1\setminus \Lambda_0}
\subset \A_{\Lambda_1}.
$$
The algebra of {\em local observables} is then defined by
$$
\A_{\rm loc} =\bigcup_{\Lambda\in\fsets} \A_\Lambda
$$
Its completion is the $C^*$-algebra of {\em quasi-local observables}, which we
will simply denote by $\A$.

The dynamics and symmetries of a quantum spin model are described by (groups
of) automorphisms of the $C^*$-algebra $\A$, i.e., bijective linear
transformations  $\alpha$ on $\A$ that preserve the product and $^*$
operations. E.g., the translation automorphisms $\tau_x$, $x\in \Ir^d$, which
map any subalgebra $\A_\Lambda$ to $\A_{\Lambda +x}$, in the natural way,
form a representation of the additive group $\Ir^d$ on $\A$.

A translation invariant {\em interaction}, or {\em potential}, defining a
quantum spin model, is a map $\phi:\fsets\to\A$ with the following
properties: for all $X\in\fsets$, we have $\phi(X)\in \A_X$,
$\phi(X)=\phi(X)^*$, and for $x\in \Ir^d$, $\phi(X+x)=\tau_x(\phi(X))$. An
interaction is called {\em finite range} if there exists $R>0$ such that
$\phi(X)=0$ whenever ${\rm diam}(X)>R$. The {\em Hamiltonian} in $\Lambda$ is
the self-adjoint element of $\A_\Lambda$ defined by 
$$
H_\Lambda=\sum_{X\subset\Lambda} \phi(X) 
$$ 
E.g., the Heisenberg model has 
\be 
\phi(\{x,y\})=- J \bS_x\cdot\bS_y, \mbox{ if } \vert x-y \vert =1,
\label{heisenberg_interaction}\ee 
and $\phi(X)=0$ in all other cases. Here,
$\bS_x\cdot\bS_y$ is the conventional notation for
$S^1_xS^1_y+S^2_xS^2_y+S^3_xS^3_y$. The magnitude of the coupling constant
$J$ sets a natural unit of energy and is irrelevant from the mathematical
point of view. Its sign, however, determines whether the model is {\em
ferromagnetic} ($J>0$), or {\em antiferromagnetic} ($J<0$). For the classical
Heisenberg model, where the role of $\bS_x$ is played by a  unit vector in
$\Rl^3$, and which can be regarded, after rescaling by a factor $S^{-2}$, as
the limit $S\to \infty$ of the quantum Heisenberg model, there is a simple
transformation relating the ferro- and antiferromagnetic models (just map
$\bS_x$ to $-\bS_x$ for all $x$ in the even sublattice of $\Ir^d$). It is
easy to see that there does not exist an automorphism of $\A$ mapping $\bS_x$
to $-\bS_x$, since that would be inconsistent with the commutation relations
\eq{commutation}. Not only is there no exact mapping between the ferro- and
the anitferromagnetic models, their ground states and equilibrium states have
radically different  properties. See below for the definitions and further
discussion. 

The {\em dynamics} (or {\em time evolution}), of the system in finite volume
$\Lambda$ is the one-parameter group of automorphisms of $\A_\Lambda$ given
by,  
$$ 
\alpha^{(\Lambda)}_t(A)=e^{itH_\Lambda} A e^{-itH_\Lambda},\quad
t\in\Rl. 
$$ 
For each $t\in \Rl$, $\alpha^{(\Lambda)}_t$ is an automorphism of
$\A$ and the family $\{\alpha^{(\Lambda)}_t\mid t\in\Rl\}$, forms a
representation of the  additive group $\Rl$.

Each $\alpha^{(\Lambda)}_t$ can trivially be extended to an automorphism on 
$\A$, by tensoring with the identity map. 
Under quite general conditions, $\alpha^{(\Lambda)}_t$
converges strongly as $\Lambda \to \Ir^d$ in a suitable sense, i.e., 
for every $A\in\A$, the limit
$$
\lim_{\Lambda\uparrow\Ir^d}\alpha^{(\Lambda)}_t(A)=\alpha_t(A)
$$
exists in the norm in $\A$, and it can be shown that it defines a 
strongly continuous one-parameter group of automorphism of $\A$.
$\Lambda\uparrow\Ir^d$ stands for any sequence of $\Lambda\in \fsets$ such
that $\Lambda$ eventually contains any given element of $\fsets$.
A sufficient condition on the potential $\phi$ is that there exists
$\lambda>0$ such that $\Vert\Phi \Vert_\lambda$ is finite, with
\be
\Vert\Phi \Vert_\lambda = \sum_{X\ni 0} e^{\lambda \vert X\vert}
\Vert \phi(X)\Vert \, .
\label{normphi}\ee
Here, $\vert\cdot\vert$ denotes the number of elements in $X$.
One can show that under the same conditions, $\delta$ defined on $\A_{\rm
loc}$ by
$$
\delta (A) = \lim_{\Lambda\uparrow\Ir^d} [H_\Lambda,A]
$$
is a norm-closable (unbounded) derivation on $\A$ and that its closure 
is, up to a factor $i$, the generator of $\{\alpha_t\mid t\in\Rl\}$, 
i.e., formally
$$
\alpha_t=e^{it\delta}\, .
$$
For the class of $\phi$ with finite $\Vert \Phi \Vert_\lambda$ for some
$\lambda>0$, $\A_{\rm loc}$ is a core of analytic vectors for $\delta$.  
This means that for each $A\in\A_{\rm loc}$, the function $t\mapsto
\alpha_t(A)$, can be extended to an entire function, which will denote
by $\alpha_z(A), z\in \Cx$. 

A {\em state} of the quantum spin system is a linear functional
on $\A$ such that $\omega(A^*A)\geq 0$, for all $A\in \A$ (positivity),
and $\omega(\idty)=1$ (normalization). The restriction of $\omega$ to 
$\A_\Lambda$, for each $\Lambda\in\fsets$, is uniquely determined by a 
{\em density matrix}, i.e., $\rho_\lambda \in \A_\Lambda$, such that 
$$
\omega(A)=\Tr \rho_\Lambda A, \mbox{ for all } A\in \A_\Lambda\, ,
$$
where $\Tr$ denotes the usual trace of matrices. $\rho_\Lambda$ is 
non-negative definite and of unit trace. If the density matrix is
a one-dimensional projection, the state is called a {\em vector state},
and can be identified with a vector $\psi\in\H_{\Lambda}$, such
that $\Cx\psi= \ran \rho_\Lambda$.

A {\em ground state} of the quantum spin system is a state $\omega$
satisfying the {\em local stability} inequalties:
\be
\omega(A^*\delta(A))\geq 0, \mbox{ for all } A\in A_{\rm loc}\, .
\label{gs}\ee

The states describing {\em thermal equilibrium} are charaterized by the
Kubo-Martin-Schwinger (KMS) condition: for any $\beta\geq 0$ (related to
absolute {\em temperature} by $\beta =1/(k_B T)$, where $k_B$ is the
Boltzmann constant), $\omega$ is called $\beta$-KMS if
\be
\omega(A\alpha_{i\beta}(B))=\omega(BA),\,\mbox { for all }
A,B\in\A_{\rm loc}\, . 
\label{KMS}\ee

The most common way to construct ground states and equilibrium states,
solutions of \eq{gs} and \eq{KMS} respectively, is by taking thermodynamic
limits of finite volume states with suitable boundary conditions. A ground
state of  the finite-volume Hamiltonian $H_\Lambda$, is a convex combination
of vector states that are eigenstates of $H_\Lambda$ belonging to its 
smallest eigenvalue. The finite-volume equilibrium state at inverse
temperature $\beta$ has denstity matrix $\rho_\beta$ defined by
$$
\rho_\beta=Z(\Lambda,\beta)^{-1}e^{-\beta H_\Lambda}
$$
where $Z(\Lambda,\beta)=\Tr e^{-\beta H_\Lambda}$, is called the
{\em partition function}. By considering limit points as $\Lambda\to\Ir^d$,
one can show that a quantum spin model has always at least one
ground state and at least one equilibrium state for all $\beta$.

What we have discussed in this section are the basic concepts in the most
standard setup. Clearly, many generalizations are possible: one can consider
non-translation invariant models, models with random potentials, the state
spaces at each site may have different dimensions, instead of $\Ir^d$ one can
consider other lattices or one can define models on arbitrary graphs, one can
allow interactions of infinite range that satisfy weaker conditions than those
imposed by the finiteness of the norm \eq{normphi}, one can restrict to
subspaces of the Hilbert space by imposing symmetries or suitable hardcore
conditions, and one can study models with infinite-dimensional spins. Examples
of all these types of generalizations have been considered in the literature
and have interesting applications.

\section{Symmetries and symmetry breaking}

Many interesting properties of quantum spin systems are related to symmetries
and symmetry breaking. Symmetries of a quantum spin model are realized as
representations of groups, Lie algebras, or quantum (group) algebras on the
Hilbert space and/or the observable algebra. The symmetry property of the
model is expressed by the fact that the Hamiltonian (or the dynamics)
commutes with this representation. We briefly discuss the most common
symmetries.

{\em Translation invariance.} We already defined the translation automorphisms
$\tau_x$ on the observable algebra of infinite quantum spin systems on
$\Ir^d$. One can also define translation automorphisms
for finite systems with periodic boundary conditions, i.e., defined on the
torus $\Ir^d/T\Ir^d$, where $T=(T_1,\ldots,T_d)$ is a positive integer vector
representing the periods.   

{\em Other graph automorphisms.} In general, if $G$ is a group of
automorhisms of the graph $\Gamma$, and $\H_\Gamma=\bigotimes_{x\in\Gamma}
\Cx^n$ is the Hilbert space of a system of identical spins defined on
$\Gamma$, then, for each $g\in G$, one can define a unitary $U_g$ on
$\H_\Gamma$ by linear extension of $U_g\bigotimes\varphi_x=\bigotimes
\varphi_{g^{-1}(x)}$, where $\varphi_x\in\Cx^n$, for all $x\in \Gamma$. These
unitaries form a representation of $G$. With the unitaries one can
immediately define automorphisms of the algebra of observables: for
$A\in\A_\Lambda$, and  $U\in \A_\Lambda$ unitary, $\tau(A)=U^*AU$ defines an
automorphism, and if $U_g$ is a group representation the corresponding
$\tau_g$ will be, too. Common examples of graph automorphisms are the
lattice symmetries of rotation and reflection. Translation symmetry and other
graph automorphisms are often referred to collectively as {\em spatial
symmetries}.

{\em Local symmetries} (also called {\em gauge symmetries}).  Let $G$ be a
group and $u_g, g\in G$, a unitary representation of $G$ on $\Cx^n$. Then,
$U_g =\bigotimes_{x\in \Lambda} u_g$, is a representation on $\H_\Lambda$.
E.g., the Heisenberg model \eq{heisenberg_interaction} commutes with such a
representation of $SU(2)$. It is often convenient, and generally equivalent,
to work with a representation of the Lie algebra. E.g., the 
$SU(2)$-invariance of the Heisenberg model is then expressed by the fact that
$H_\Lambda$ commutes with the following three operators: 
$$ 
S^i =\sum_{x\in\Lambda} S^i_x\, ,\quad i=1,2,3\, .
$$

Note: sometimes the Hamiltonian is only symmetric under certain
combinations of spatial and local symmetries. CP symmetry is an example.

For an automorphism $\tau$, we say that a state $\omega$ is $\tau$-invariant
if $\omega\circ\tau=\tau$. If $\omega$ is $\tau_g$-invariant
for all $g\in G$, we say that $\omega$ is $G$-invariant.

It is easy to see that if a quantum spin model has a symmetry $G$, then
the set of all ground states or all $\beta-$KMS states will be $G$-invariant,
meaning that if $\omega$ is in the set, then so is $\omega\circ\tau_g$,
for all $g\in G$. By a suitable averaging procedure it is ususally easy
to establish that the sets of ground states or equilibrium states contain
at least one $G$-invariant element.

An interesting situation occurs if the model is $G$-invariant, but
there are ground states or KMS states that are not. I.e., for some
$g\in G$, and some $\omega$ in the set (of ground states or KMS states), 
$\omega\circ\tau\neq\omega$. When this happens, one says  that there
is {\em spontaneous symmetry breaking}, a phenomenon that also plays an
important role in Quantum Field Theory.

The famous Hohenberg-Mermin-Wagner Theorem, applied to quantum spin models,
states that, as long as the interactions are not too long range and the
dimension of the lattice is two or less, continuous  symmetries cannot be
spontaneously broken in a $\beta$-KMS state for any finite $\beta$.

{\em Quantum group symmetries.}  We restrict ourselves to one important
example: the $SU_q(2)$-invariance of the spin-1/2 XXZ Heisenberg chain
with $q\in [0,1]$, and with 
special boundary terms. The Hamiltonian of the $SU_q(2)$-invariant
XXZ-chain of length $L$ is 
\beann
&&H_L=\sum_{x=1}^{L-1} 
-\frac{1}{\Delta}(S^1_x S^1_{x+1}+ S^2_x S^2_{x+1})\\
&&-(S^3_x S^3_{x+1} - 1/4) 
+ \frac{1}{2}\sqrt{1-\Delta^{-2}}(S^3_{x+1} - S^3_x)\, ,
\eeann
where $q\in (0,1]$ is related to the parameter $\Delta \geq 1$ by the
relation $\Delta=(q+q^{-1})/2$. When $q=0$, $H_l$ is equivalent to the Ising
chain. Thus, the XXZ model interpolates between the Ising model (the
primordial classical spin system) and the isotropic Heisenberg model (the
most widely studied quantum spin model). In the limit of infinite spin
($S\to\infty$), the model converges to the classical Heisenberg model (XXZ or
isotropic). An interesting feature of the XXZ model are its non translation
invariant ground states, called kink states.

In this family of models one can see how aspects of discreteness
(quantized spins) and continuous symmetry (SU(2), or quantum symmetry
$SU_q(2)$) are present at the same time in the quantum Heisenberg models,
and the two classical limits ($q\to 0$ and $S\to \infty$), can be used 
as a starting point to study its properties.

Quantum group symmetry is not a special case of invariance under the action
of a group. There is no group. But there is an algebra represented on  the
Hilbert space of each spin, for which there is a good definition of tensor
product of representations, and ``many'' irreducible representations. In this
example the representation of $SU_q(2)$ on $\H_{[1,L]}$ commuting with $H_L$
is generated by
\beann
S^3\!&=&\!\sum_{x=1}^L \idty_1\otimes\cdots\otimes
S^3_x\otimes\idty_{x+1}\otimes\cdots\idty_L\label{spinmua}\\
S^+\!&=&\!\sum_{x=1}^L t_1\otimes\cdots\otimes t_{x-1}\otimes
S^+_x\otimes\idty_{x+1}\otimes\cdots\idty_L\label{spinmub}\\
S^-\!&=&\!\sum_{x=1}^L \idty_1\otimes\cdots\otimes
S^-_x\otimes t^{-1}_{x+1}\otimes\cdots t^{-1}_L
\eeann
where $$
t=\left(\begin{array}{cc} q^{-1}&0\\0&q\end{array}\right)\, .
$$
Quantum group symmetries were discovered in exactly solvable models, starting
with the spin-1/2 XXZ chain. One can exploit their representation
theory to study the spectrum of the Hamiltonian in very much the same 
way as ordinary symmetries. The main restriction to its applicability is
that the tensor product structure of the representations is inherently
one-dimensional, i.e., relying on an ordering from left to right.
For the infinite XXZ chain the left-to-right and right-to-left orderings 
can be combined to generate an infinite-dimensional algebra, 
the quantum affine algebra $U_q(\widehat{sl}_2)$.

\section{Phase Transitions}

Quantum spin models of condensed matter physics often have interesting ground
states. Not only are the ground states often a good approximation of the
low-temperature behavior of the real systems that are modeled by it, and
studying them is therefore useful, it is in many cases also a challenging
mathematical problem. This is in contrast with classical lattice models for
which the ground states are ususally simple and easy to find. In more than one
way ground states of quantum spin systems display behavior similar to 
equilibrium states of classical spin systems at positive temperature.

The spin-1/2
Heisenberg antiferromagnet on $\Lambda\subset \Ir^d$, with Hamiltonian
\be
H_\Lambda = \sum_{x,y;\in\Lambda\vert x-y\vert=1}
\bS_x\cdot \bS_y\, ,
\ee
is a case in point. Even in the one-dimensional case ($d=1$), and even though
the model in that case is exactly solvable by the Bethe Ansatz, its ground 
state is highly non-trivial. Analysis of the Bethe Ansatz solution (which is
not fully rigorous) shows that spin-spin correlation function decays to
zero at infinity, but slower than exponentially (roughly as inverse distance
squared). For $d=2$, it is believed but not mathematically proved that
the ground state has N\'eel order, i.e., long-range antiferromagnetic order,
accompanied by a spontaneous breaking of the SU(2) symmetry. Using {\em
reflection positivity}, Dyson, Lieb, and Simon were able to prove N\'eel
order at  sufficiently low temperature (large $\beta$), for $d\geq 3$ and all
$S\geq 1/2$. This was later extended to the ground state for $d= 2$
and $S\geq 1$, and $d\geq 3$ and $S\geq 1/2$, i.e., all cases where 
N\'eel order is expected  except $d=2$, $S=1/2$.

In contrast, no proof of long range order in the Heisenberg ferromagnet at low
temperature exists. This is rather remarkable since proving long range order
in the ground states of the ferromagnet is a trivial problem.

Of particular interest are the so-called quantum phase transitions. These are
phase transitions that occur as a parameter in the Hamiltonian is varied and
which are driven by the competing effects of energy and quantum fluctuations,
rather than the balance between energy and entropy which drives usual
equilibrium phase transitions. Since entropy does not play a role, quantum
phase transitions can be oberved at zero temperature, i.e., in the ground
states.

An important example of a quantum phase transition occurs in the two- or
higher dimensional $XY$-model with a magnetic field in the  $Z$-direction. It
was proved by Kennedy, Lieb, and Shastry that, at zero field, this model has
Off-Diagonal-Long-Range-Order (ODLRO), and can be interpreted as a hard-core
bose gas at half-filling. It is also clear that if the magnetic field exceeds
a critical value, $h_c$, the model has a simple ferromagnetically ordered
ground state. There are indications that there is ODLRO for all $\vert h\vert
<h_c$. However, so far there is no proof that ODLRO exists for any $h\neq 0$.

What makes the ground state problem of quantum spin systems interesting and
difficult at the same time is that ground states, in general, do not
minimize the expectation of the interaction terms in the Hamiltonian
individually although, loosely speaking, the expectation of their sum (the
Hamiltonian) is minimized. However, there are interesting exceptions to
this rule. Two examples are the AKLT model and the ferromagnetic XXZ model.

The wide ranging behavior of quantum spin models has required an equally wide
range of mathematical approaches to study them. There is one group of methods,
however, that can make a claim of substantial generality: those that start
from a representation of the partition function based on the Feynman-Kac
formula. Such representations turn a $d$-dimensional quantum spin model into a
$d+1$-dimensional classical problem, albeit one with some special features.
This technique was pioneered by Ginibre in 1968 and was quickly
adopted by a number of authors to solve a variety of problems.
Techniques borrowed from classical statistical mechanics have been adapted
with great success to study ground states, the low-temperature phase diagram,
or the high-temperature regime of quantum spin models that can be regarded as
perturbations of a classical system.  More recently, it was used to develop a
quantum version of Pirogov-Sinai theory which is applicable to a large class 
of problems, including some with low-temperature phases not related by
symmetry.

\section{Dynamics}

Another feature of quantum spin systems that makes them mathematically richer
than their classical couterpart, is the existence of a Hamiltonian dynamics.
We have seen that, quite generally, the dynamics is well-defined in the
thermodynamic limit as a strongly continuous one-parameter group of automorphisms
of the $C^*$-algebra of quasi-local observables.  Strictly speaking, a
quantum spin model is actually {\em defined} by its dynamics $\alpha_t$, or
by its generator $\delta$, and not by the potential $\phi$. Indeed,  $\phi$
is not uniquely determined by $\alpha_t$. In particular, it is possible to
incorporate various types of boundary condition into the definition of
$\phi$.  This approach has proved very useful in obtaining important
structural results, such as, e.g., the proof by Araki of the uniqueness the
KMS state at any finite $\beta$ in one-dimension.  Another example is a
characterization of equilibrium states  by the Energy-Entropy Balance
inequalities, which is both  physically appealing and mathematically useful:
$\omega$ is a $\beta$-KMS state for a quantum spin model in the setting of
section \ref{sec:frame} (and in fact also for more general quantum systems),
if and only if the inequality 
$$
\beta\omega(X^*\delta(X))\geq\omega(X^*X)\log\frac{\omega(X^*X)}{\omega(XX^*)}
$$ 
is satisfied for all $X\in\A_{\rm loc}$. This characterization and
several related results were proved in a series works by various authors
(mainly Roepstorff, Araki, Fannes, Verbeure, and Sewell).

Detailed properties of the dynamics for specific models are generally lacking.
One could point to the ``immediate non-locality'' of the dynamics as the main
difficulty. By this, we mean that, except in trivial cases, most
local observables $A\in\A_{\rm loc}$, become non-local after an arbitrarily
short time, i.e., $\alpha_t(A)\not\in \A_{\rm loc}$, for any $t\neq 0$.
This non-locality is not totally uncontroled however. A result by
Lieb and Robinson establishes that, for models with interactions that
are sufficiently short range (e.g., finite range), the non-locality 
propagates at a bounded speed. More precisely, under quite general
conditions, there exists constants, $c,v>0$, such that for any two local 
observables $A, B \in \A_{\{0\}}$, 
$$
\Vert [\alpha_t(A),\tau_x(B)]\Vert \leq 2\Vert A\Vert \Vert B\Vert
e^{-c(\vert x \vert-v\vert t\vert)}\, .
$$

Attempts to understand the dynamics have generally been aimed at one of  two
issues: return to equilibrium from a perturbed state, and convergence  to a
non-equilibrium steady state in the presence of currents. Some interesting
results have been obtained although much remains to be done.

\section*{See Also}

Phase transitions. $C^*$-algebras. UHF-algebra. Quantum phase transitions.
Reflection positivity. Hubbard model. Heisenberg model. Bethe Ansatz.
Falicov-Kimball model. Symmetry breaking. Finitely Correlated States. XY model.
Thermodynamic limit. Magnetism. Quantum information. $SU(2)$. $SU_q(2)$.
$U_q(\widehat{sl}_2)$. Density Matrix Renormalization Group.
Hohenberg-Mermin-Wagner Theorem. Integrable spin chains.

\section*{Further Reading}

A very informative overview of the early history of quantum spin systems,
especially in relation to the history of the theory of magnetism and
including a good bibliography, can be found in \cite{Mattis}. 

The mathematical framework briefly described in Section \ref{sec:frame} is 
discussed in detail in \cite{BR2}. 

\cite{KN} is an annotated bibliography devoted to mathematical results
for the Heisenberg and related models.

Since many important results and
techniques have not yet appeared in book form, we have included some 
seminal research papers of the field in the references.

\providecommand{\bysame}{\leavevmode\hbox to3em{\hrulefill}\thinspace}

\section*{Keywords}

spin matrices, Heisenberg model, phase transitions, quantum phase
transitions, Bethe Ansatz, spontaneous symmetry breaking, ground state, KMS
state, equilibrium state, Pauli matrices, $SU(2)$, $SU_q(2)$, spin waves,
reflection positivity, $C^*$-algebra, N\'eel order, Pirogov-Sinai theory,
quantum Pirogov-Sinai theory, magnetism, ferromagnet, antiferromagnet,
long range order, off-diagonal long range order, kink state, XXZ model,
entropy-energy inequalities, quantum spin dynamics

\end{document}